\documentclass{article}

\usepackage{graphics}
\usepackage{emulateapj}

\newbox\grsign \setbox\grsign=\hbox{$>$} \newdimen\grdimen \grdimen=\ht\grsign
\newbox\simlessbox \newbox\simgreatbox
\setbox\simgreatbox=\hbox{\raise.5ex\hbox{$>$}\llap
     {\lower.5ex\hbox{$\sim$}}}\ht1=\grdimen\dp1=0pt
\setbox\simlessbox=\hbox{\raise.5ex\hbox{$<$}\llap
     {\lower.5ex\hbox{$\sim$}}}\ht2=\grdimen\dp2=0pt
\def\simgreat{\mathrel{\copy\simgreatbox}}
\def\simless{\mathrel{\copy\simlessbox}}

\newcommand{\hhh}{h_{100}}
\newcommand{\m}{$^{-1}$}
\newcommand{\ROSAT}{\emph{ROSAT}}
\newcommand{\ASCA}{\emph{ASCA}}
\newcommand{\grad}{\nabla}
\newcommand{\mdot}{\dot{M}}

\newcommand{\dlndln}[2]{\frac{d \, {\ln{#1}}}{d \, {\ln{#2}}}}
\newcommand{\dd}[2]{\frac{d {#1}}{d {#2}}}
\newcommand{\mr}[1]{\mathrm{#1}}
\newcommand{\btrue}{\beta_\mr{true}}
\newcommand{\bfit}{\beta}
\newcommand{\betamodel}{$\bfit$-model}
\newcommand{\rhodm}{\rho_\mr{dm}}

\newcommand{\etal}{et al.\ }

\begin{document}
\submitted{Submitted April 11, 2000; Accepted August 2, 2000 for 
publication in the \emph{Astrophysical Journal}}
\title{The Orbital Structure of Dark Matter Halos With Gas}

\author{Andisheh Mahdavi}
\affil{Harvard-Smithsonian Center for Astrophysics, MS 10, 60
Garden St., Cambridge, MA 02138; amahdavi@cfa.harvard.edu}

\begin{abstract}
With the success of the \emph{Chandra} and \emph{XMM} missions and the
maturation of gravitational lensing techniques, powerful constraints
on the orbital structure of cluster dark matter halos are possible. I
show that the X-ray emissivity and mass of a galaxy cluster uniquely
specify the anisotropy and velocity dispersion profiles of its dark
matter halo. I consider hydrostatic as well as cooling flow scenarios,
and apply the formalism to the lensing cluster CL0024+16 and the
cooling flow cluster Abell 2199. In both cases, the model predicts a
parameter-free velocity dispersion profile that is consistent with
independent optical redshift surveys of the clusters.
\end{abstract}


\section{Introduction}

In the standard picture, a cluster of galaxies consists of a dark
matter halo accompanied by galaxies and a hot, 1-15 keV plasma. The
techniques for measuring the mass of the halo usually focus
exclusively on the motions of the galaxies in the cluster, on the
emissivity and temperature of the X-ray emitting gas, or on the
gravitational lensing of distant sources by the halo. In this paper I
show that combining these various data can place strong constraints on
the velocity structure of the cluster dark matter halo.

Over the past two decades, a great deal of evidence has mounted that
the X-ray emissivity $\epsilon$ of the plasma for many clusters is
well fit by the particularly simple function (e.g. Jones \& Forman
1984; Mohr, Mathiesen, \& Evrard 1999)
\begin{equation}
\epsilon \propto \left(1 + \frac{r^2}{r_c^2} \right)^{-3 \bfit}.
\label{eq:betamodel}
\end{equation}
This function is often referred to as the \betamodel\ in the
literature; it has a core radius, $r_c$, and a faint end slope,
$\bfit$. 

It is instructive to review the physical basis for the form of the
\betamodel\ (see, e.g. Sarazin 1988). The starting point is the
assumption that the plasma is in equilibrium with the dark matter
potential. The general form of the equation of hydrostatic equilibrium
is
\begin{equation}
\grad P = - \rho_g \grad \phi,
\end{equation}
where $P$ is the gas pressure, $\rho_g$ is the gas density, and $\phi$
is the gravitational potential. With the assumption that the plasma is
a spherically symmetric ideal gas, the equation becomes
\begin{equation}
\frac{1}{\rho_g} 
\dd{}{r} \left( \frac{\rho_g k_B T}{m_p \mu} \right) = -\frac{G M}{r^2} 
\label{eq:bhydro}
\end{equation}
where $T$ is the plasma temperature, $M$ is the total mass inside the
radius $r$, $m_p$ is the proton mass, $\mu$ is the mean molecular
weight, and $k_B$ is Boltzmann's constant. The dynamics of the dark
matter halo, on the other hand, are determined by the Jeans equation
(e.g. Binney \& Tremaine 1987, p. 204),
\begin{equation}
\frac{1}{\rhodm} \dd{}{r} \left(\rhodm \sigma_r^2\right) + 
\frac{2 \eta \sigma_r^2}{r} = - \frac{G M}{r^2}.
\label{eq:bjeans}
\end{equation}
Here $4 \pi r^2 \rhodm = d M / dr$ is the dark matter density,
$\sigma_r$ is the dark matter radial velocity dispersion, and $\eta$
is the anisotropy parameter\footnote{ In the stellar dynamics
literature, which precedes the discussion of X-ray clusters by many
decades, it is $\beta$ that denotes the anisotropy parameter. But
because the $\beta$ is so widely used by X-ray astronomers to refer to
the faint-end slope of the emissivity, I must switch symbols.}, which
 describes, in an average sense, the nature of the orbits of the dark
matter particles. The anisotropy parameter is a dimensionless quantity
equal to $1 - \sigma_\theta^2 / \sigma_r^2$, where $\sigma_\theta^2$
is the tangential velocity dispersion. When $\eta = 1$, the dark
matter orbits are completely radial; when $\eta = -\infty$, the orbits
are completely tangential; and when $\eta = 0$, the orbits are
isotropic.  Note that equation (\ref{eq:bjeans}) is not the same as
the galaxy Jeans equation, which involves the galaxy number density
$\nu_\mr{gal}$ in place of the dark matter density $\rhodm$.

Three critical assumptions shape the derivation of the \betamodel.
The first is that the gas is isothermal, and the second is that the
dark matter orbits are perfectly isotropic. Then, equating
(\ref{eq:bhydro}) and (\ref{eq:bjeans}),
\begin{equation}
\dlndln{\rho_g}{r} = \dlndln{\rhodm}{r}  \frac{\mu m_p \sigma_r^2}{k_B T}.
\end{equation}
The final assumption is that the dark matter halo is nearly an
isothermal sphere, with a density profile given by a King model,
$\rhodm \propto (1 + r^2/r_c^2)^{-3/2}$. Then the above equation has
$\rho_g \propto (1 + r^2/r_c^2)^{- 3 \beta / 2}$, with $\beta \equiv \mu
m_p \sigma_r^2 / (k_B T)$. The isothermal emissivity,
proportional to the square of the gas density, is then given by 
equation (\ref{eq:betamodel}).

There is an inconsistency in the above derivation. The assumption that
$\eta = 0$ everywhere is not compatible with a constant velocity
dispersion $\sigma_r$, easily verified by inserting a King model into
equation (\ref{eq:bjeans}), setting $\eta = 0$, and noticing that
$\sigma_r$ must vary with the radius. Yet the \betamodel\ derived in
this manner continues to provide an adequate empirical description of
cluster emissivities.

It is interesting to ask what anisotropy profiles $\eta(r)$ and
velocity dispersion profiles $\sigma_r^2(r)$ are fully consistent with
the \betamodel\ taken in conjunction with auxiliary measurements,
e.g. gravitational lensing. A further impetus for investigating $\eta$
is the recent development (e.g. Geller, Diaferio, \& Kurtz 1999) that
the density profiles of some clusters of galaxies are well fit by a
cuspy model suggested by Navarro, Frenk, \& White (1997; NFW); see
Table \ref{tbl:potentials}. Because gasdynamical N-body simulations
indicate that a \betamodel\ gas can coexist with an NFW dark halo
(e.g. Eke \etal 1998), an analytic exploration of the basic
equilibrium properties of the anisotropy parameter is
appropriate. Finally, a theoretically sound functional form for $\eta$
is required for analyzing cluster velocity dispersion profiles using
the Jeans equations (e.g., Carlberg \etal 1997).

In this paper I develop a formalism for calculating $\eta(r)$ and
$\sigma_r^2(r)$ simultaneously. I assume that the cluster is
spherically symmetric, that the gas is in hydrostatic or
quasihydrostatic equilibrium with the dark matter, that dark matter
dominates the cluster potential, and that the gas temperature is
always proportional to the local dark matter velocity dispersion. In
\S \ref{sec:derive} I derive the basic equations; in \S
\ref{sec:hydro}, I apply the model to hydrostatic plasmas and the
cluster CL0024+16; in \S \ref{sec:cooling}, I consider anisotropy
profiles within quasihydrostatic cooling flows, specifically the
cluster Abell 2199; and in \S \ref{sec:conclusion} I summarize.

\section{Derivation}
\label{sec:derive}

The aim of this derivation is to solve the dark matter Jeans
equation (\ref{eq:bjeans}) for the radial velocity dispersion
$\sigma_r$ and the orbital anisotropy parameter $\eta$ given a mass
profile $M$.  Because there are two unknown functions and only one
equation, auxiliary assumptions are necessary.  One approach is to
make $\eta$ a constant (Mahdavi \etal 1999; van der Marle 2000) and
solve for $\sigma_r(r,\eta)$. However, N-body simulations suggest that
the orbital distribution of the dark matter particles varies
considerably throughout the halo (e.g. Kaufmann \etal 1999; Diaferio
2000). Another method involves assuming a functional form for the
phase space distribution of the dark matter particles (Merritt 1985;
Gerhard \etal 1998). This approach is tantamount to fixing the form of
$\eta(r)$.

Here I outline a third approach that places no requirements on $\eta$
itself. Instead I assume that the total specific energy of the dark
matter particles is everywhere proportional to that of the plasma: $T
\propto \sigma_\mr{tot}^2$. This scaling law is not arbitrary, but is
grounded in observations of groups and clusters of galaxies. Combined
X-ray and optical observations of these systems have found that the
mean emission-weighted gas temperature scales roughly as the second
power of the total galaxy velocity dispersion (Mulchaey \& Zabludoff
1998; Xue \& Wu 2000). These data show that in many clusters, the dark
matter, gas, and the galaxies are in dynamical equilibrium, with the
specific energy of each component proportional to that of the other
two at all observable radii.

Now I describe the implications of $T \propto \sigma_\mr{tot}^2$ for
the anisotropy parameter. In a spherically symmetric dark matter halo,
the velocity dispersion tensor is close to diagonal in spherical
coordinates. The local energy per unit mass is therefore given by the
trace of the tensor, the total three-dimensional velocity dispersion:
\begin{eqnarray}
\sigma_\mr{tot}^2 & = & \sigma_r^2 + \sigma_\theta^2 + \sigma_\phi^2 \\
                  & = & \sigma_r^2 \left( 3 - 2 \eta \right),
\end{eqnarray}
where $\sigma_\theta$ and $\sigma_\phi$ are the tangential and
azimuthal velocity dispersions, respectively. In spherically symmetric
systems, $\sigma_\theta = \sigma_\phi$ and and $\eta = 1 -
\sigma_\theta^2/\sigma_r^2$. Then the hypothesis that $T \propto
\sigma_\mr{tot}^2$ implies the following relationship between the gas
temperature and the velocity dispersion:
\begin{equation}
3 \btrue k_B T  =  \mu m_p \sigma^2_r \left(3 - 2 \eta \right),
\label{eq:btrue}
\end{equation}
where $\btrue$ is a dimensionless constant.

Equations (\ref{eq:bhydro}), (\ref{eq:bjeans}), and (\ref{eq:btrue})
now form a closed set, solvable as follows. I begin by
writing the equation of hydrostatic equilibrium in a different form:
\begin{equation}
\dlndln{\rho_g T}{r} = -\frac{G M m_p \mu}{k_B T r} 
\label{eq:hydro}
\end{equation}

In clusters of galaxies, the plasma radiates chiefly through thermal
bremsstrahlung, a collisional process whose total emissivity scales
roughly as the square root of the temperature: $\epsilon \propto
\rho_g^2 \sqrt{T}$. It follows that
\begin{equation}
\ln{\rho_g} = \frac{1}{2}\ln{\epsilon} - \frac{1}{4} \ln{T} + \mr{constant}.
\end{equation}
It is then possible to reformulate equation (\ref{eq:hydro})  in
terms of the observed emissivity:
\begin{equation}
\frac{3}{4} \dlndln{T}{r} + \frac{1}{2} \dlndln{\epsilon}{r} =
 -\frac{G M m_p \mu}{k_B T r}.
\label{eq:dlntdr}
\end{equation}
This relation implies that, for isothermal plasmas, there is a
one-to-one correspondence between the measured emissivity profile and
the mass profile. In other words, an isothermal sphere of ideal gas
with a measured emissivity profile $\epsilon(r)$ must have the mass
profile given by setting $d \ln{T} / d \ln{r} = 0$ in equation
(\ref{eq:dlntdr}):
\begin{equation}
M(r) = -\frac{k_B T r}{2 G m_p \mu} \dlndln{\epsilon}{r}.
\label{eq:iso}
\end{equation}

In general, the plasma can have a temperature gradient; the profile is
given by the solution of equation (\ref{eq:dlntdr}):
\begin{equation}
T(r) = \frac{4 \mu m_p}{3 k_B \epsilon^{2/3}} \int_r^\infty
\frac{G M}{r^2} \epsilon^{2/3} d r.
\label{eq:tofr}
\end{equation}
Here the constant of integration is chosen such that $T$ converges as
$r \rightarrow \infty$. The temperature at infinity is
\begin{equation}
T_\infty = - \lim_{r \rightarrow \infty} \frac{2 G M m_p \mu}{k_B r} 
\left( \dlndln{\epsilon}{r} \right)^{-1}.
\end{equation}
If the logarithmic derivative of the emissivity tends to a constant
value, then as $r \rightarrow \infty$, (1) $T_\infty = 0$ unless $M(r)
\propto r$, and (2) $M(r) \propto r$ is the steepest mass profile
allowed. A further useful result is obtained by applying l'H\^optial's
rule to the derivative of equation \ref{eq:tofr}:
\begin{equation}
\lim_{r \rightarrow \infty} \dlndln{T}{r} = \lim_{r \rightarrow
\infty} \dlndln{M}{r} - 1.
\end{equation}
That is, $T(r) \propto M(r)/r$ asymptotically. Because the mass can
never decrease, the temperature never falls faster than $r^{-1}$ at
large radii.



Now I rewrite the Jeans equation for a spherically symmetric system of
collisionless particles:
\begin{equation}
\dlndln{\rhodm \sigma_r^2}{r} + 
2 \eta = - \frac{G M}{\sigma_r^2 r},
\label{eq:jeans}
\end{equation}
 Using equation (\ref{eq:btrue}), it is possible to eliminate $\eta$:
\begin{equation}
\dlndln{\rhodm \sigma_r^2}{r} + 3 \left(1 - 
\frac{\btrue k_B T}{m_p \mu \sigma_r^2} \right) =
 - \frac{G M}{\sigma_r^2 r}.
\end{equation}
This equation has the solution
\begin{eqnarray}
\label{eq:sigma}
\sigma_r^2(r) & = &  3 \btrue \sigma_1^2 - \sigma_2^2,  \\
\label{eq:sigone}
\sigma_1^2 & = & \frac{1}{r^3 \rhodm} \int
\frac{ k_B T}{m_p \mu} r^2 \rhodm dr, \\ 
\label{eq:sigtwo}
\sigma_2^2 & = & \frac{1}{r^3 \rhodm} 
\int G M r \rhodm   dr.
\end{eqnarray}
The constant of integration should be zero in most cases, because a
nonzero constant will usually introduce a $( r^3 \rhodm )^{-1}$
divergence in the velocity dispersion as $r \rightarrow 0$. Such a
divergence cannot be reconciled with equation (\ref{eq:btrue}) if the
temperature is finite at $r=0$ and $\eta < 1$.

Two physical considerations limit the possible values of $\btrue$. The
first is that $\sigma_r^2 \ge 0$. Then
\begin{equation}
 \sigma_2^2 \le 3 \btrue \sigma_1^2.
\label{eq:bone}
\end{equation}
The second consideration is $\eta < 1$, equivalent to requiring that
the tangential velocity dispersion $\sigma_\theta^2$ always be
positive. Equation (\ref{eq:btrue}) then implies
\begin{eqnarray}
3 \btrue k_B T & \ge & m_p \mu \sigma_r^2 \\
3 k_B T \btrue  & \ge & m_p \mu 
\left( 3 \btrue \sigma_1^2 - \sigma_2^2 \right) \\
\sigma_2^2 & \ge & 
3 \btrue \left( \sigma_1^2 - \frac{k_B T}{m_p \mu} \right)
\label{eq:btwo}
\end{eqnarray}
Any physically acceptable emissivity-potential pair must
simultaneously meet the requirements of inequalities (\ref{eq:bone})
and (\ref{eq:btwo}) at all radii. Only values of $\btrue$ which meet
these two constraints are valid. If at any two radii the two
inequalities are in conflict, the emissivity-potential pair is not
physically consistent with the assumptions of this
derivation. Whenever $T \rightarrow 0$ asymptotically and
$\sigma_{1,2}^2$ do not, the above inequalities force a unique value
of $\btrue$:
\begin{equation}
\btrue = \lim_{r \rightarrow \infty} \frac{\sigma_2^2}{3 \sigma_1^2}.
\label{eq:bliminf}
\end{equation}


Finally, for a valid choice of $\btrue$, the anisotropy profile is given
by the inversion of equation (\ref{eq:btrue}):
\begin{equation}
\eta(r) = \frac{3}{2} \left(1 - \frac{\btrue k_B T}{m_p \mu \sigma_r^2}
\right).
\label{eq:eta}
\end{equation}
Thus only the shapes, and not the normalizations, of the gravitational
potential and the emissivity profile determine the velocity
anisotropy.

\section{Hydrostatic Solutions}
\label{sec:hydro}

Here I discuss the anisotropy solutions for clusters in perfect
hydrostatic equilibrium---i.e., assuming radiative cooling is
unimportant. 
I begin with potential-emissivity pairs described by simple power
laws. Although single power laws do not provide a satisfactory
description of real clusters, they elucidate the general properties of
the solutions more clearly than broken power laws, for which the
integrals in equations (\ref{eq:sigone})-(\ref{eq:sigtwo})
are often not expressible in terms of elementary functions. 


\subsection{Power Law Mass and Emissivity Profiles}
\label{sec:powerlaw}

Consider a cluster with mass profile $M = A r^\alpha$ (with $\rhodm
\propto r^{\alpha-3}$), and  a power law emissivity profile
$\epsilon \propto r^\nu$. The mass must never decrease with the radius,
so $\alpha > 0$. Then equation (\ref{eq:tofr}) has the form
\begin{equation}
\frac{k T(r)}{m_p \mu}
  = \frac{4 G A r^{\alpha-1}}{3 (1 - \alpha - 2 \nu / 3)}.
\label{eq:powert}
\end{equation}
It is unphysical for the temperature to become infinite at large
radius. Furthermore, the temperature must always be positive. Hence we
have the constraints
\begin{eqnarray}
0  < \alpha \le  1 \\
\nu  <  \frac{3 - 3\alpha}{2}
\end{eqnarray}
Thus, if the matter distribution behaves like a singular isothermal
sphere ($\alpha = 1$), the plasma is isothermal as long as its
emissivity is a declining power law of arbitrary index.

The radial velocity dispersion from equation (\ref{eq:sigma})
is
\begin{eqnarray}
\sigma_r^2(r) &    =    & \frac{G A r^{\alpha-1}}{2 \alpha - 1} \left(
		       \frac{4 \btrue }{\gamma} - 1 \right), \\
\label{eq:powersig}
\gamma &  \equiv & 1 - \alpha - 2 \nu / 3. 
\end{eqnarray}
Again, for a singular isothermal sphere the radial velocity dispersion
is constant regardless of the power law index of the emissivity.  The
logarithmic $\alpha = 1/2$ solution is not permitted, because it gives
$\eta = 3/2$ at $r = 0$.

Substituting equations (\ref{eq:powert}) and (\ref{eq:powersig}) into
the anisotropy equation (\ref{eq:eta}) yields,
\begin{equation}
\eta = \frac{1}{2} \left(3 - \frac{8 \alpha - 4}
{4 -  \gamma / \btrue} \right).
\label{eq:anisotropy}
\end{equation}
Thus the anisotropy parameter $\eta$ is always constant for the
permitted power law mass and emissivity profiles.  Note that the dark
matter orbits are perfectly isotropic ($\eta = 0$) only if the ratio
of the dark matter velocity dispersion to the gas temperature is
\begin{equation}
\btrue = \frac{3 \gamma}{16 - 8 \alpha}.
\end{equation}

Not all values of $\btrue$ are allowed. The requirements that
$\sigma_r^2 > 0$ and $\eta < 1$ place the following bounds on
$\btrue$:
\begin{eqnarray}
\frac{\gamma}{8 - 8 \alpha}  < \btrue <  \frac{\gamma}{4} 
& \rm{\ if\ } \alpha < 1/2; \\
\frac{\gamma}{4}  < \btrue <  \frac{\gamma}{8 - 8 \alpha} 
& \rm{\ if\ } \alpha > 1/2.
\end{eqnarray}

\begin{figure*}
\resizebox{7in}{!}{\includegraphics{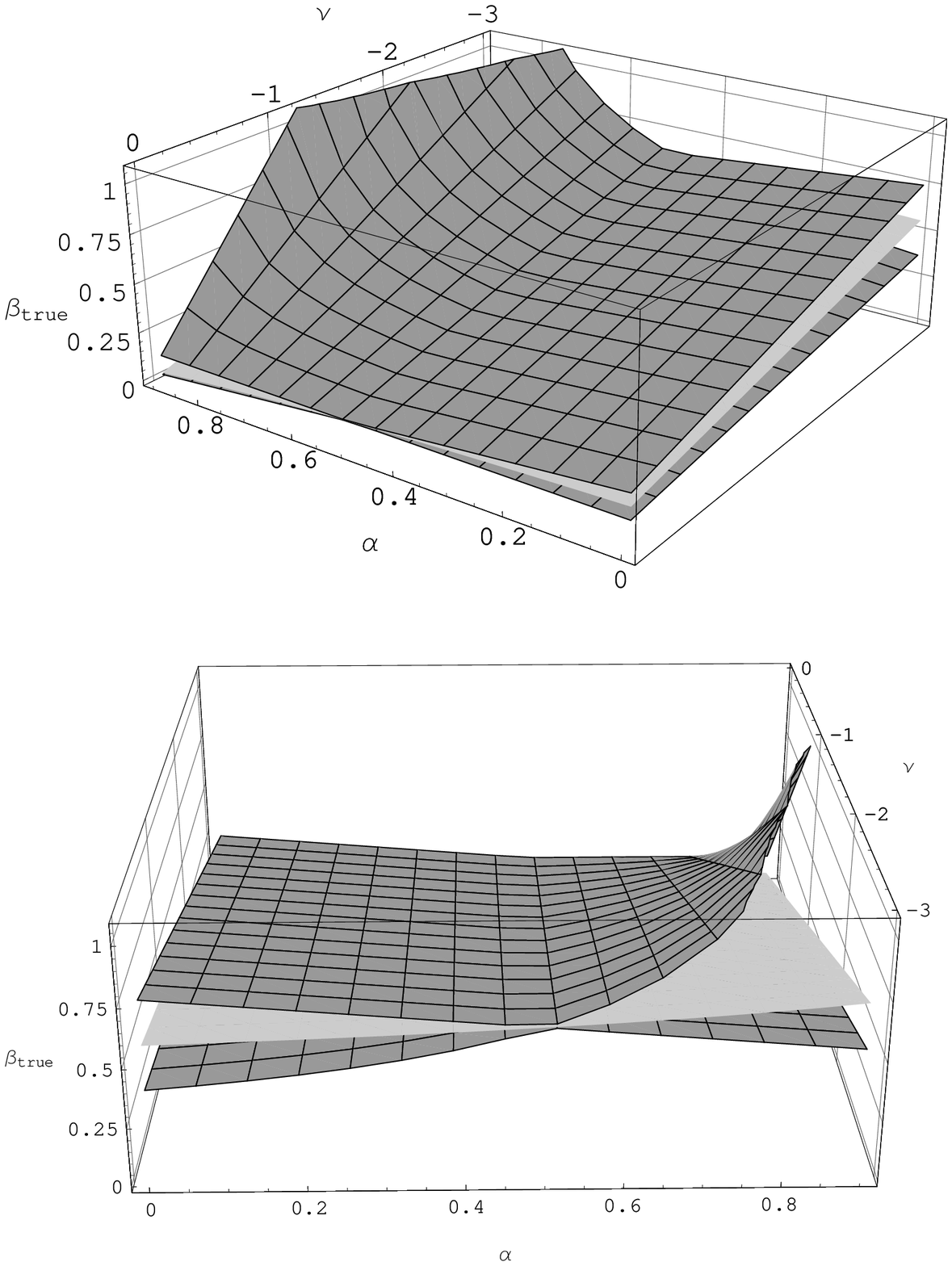}}
\figcaption[powerlaw.eps]
{Limits on $\btrue$, the ratio of the total dark matter
velocity dispersion to the gas temperature, for mass and emissivity
profiles $M \propto r^\alpha$ and $\epsilon \propto r^\nu$. Two
different views of the same plot are shown for clarity. The meshed
surfaces indicate the upper and lower limits on $\btrue$. The smooth
surface shows the required value of $\btrue$ such that the velocity
anisotropy $\eta = 0$. \label{fig:powerlaw}}
\end{figure*}

Figure \ref{fig:powerlaw} depicts these boundaries, as well as
the $\eta = 0$ criterion. For $\alpha < 1$, $\btrue$ always has
firm upper and lower limits, but for the singular isothermal
sphere, arbitrary large values of $\btrue$ are allowed. Within
the upper and lower limits, the anisotropy $\eta$ takes on the
full range of allowed values, $(-\infty,1)$.

A physical explanation of overall tilt of the planes in Figure
\ref{fig:powerlaw} lies in the slope of the emissivity profile. When
the gas distribution is relatively extended ($\nu \simgreat -1$), the
gas is on the average hotter, and the dark matter to gas energy ratio
$\btrue$ is smaller, than when the gas is relatively concentrated
($\nu < -1$). Thus the mean permitted value for $\btrue$ increases as
the emissivity profile steepens.

The asymmetry in Figure \ref{fig:powerlaw} between the boundaries for
$\alpha < 1/2$ and $\alpha > 1/2$ is related to the differences in the
circular velocity profile of the potential, $v_c = \sqrt{G M / r}$.
In systems with $\alpha > 1/2$, $v_c$ is nearly constant, and ample
kinetic energy is available at all radii to particles with tangential
orbits. It is therefore statistically difficult to underpopulate the
tangential orbits and achieve $\eta \approx 1$. Thus $\btrue$ must be
inordinately large to achieve a significant radial anisotropy: a large
amount of kinetic energy must reside in the radial orbits for them to
dominate the total velocity dispersion.



\vspace{0.1in}
\resizebox{3.5in}{!}{\includegraphics{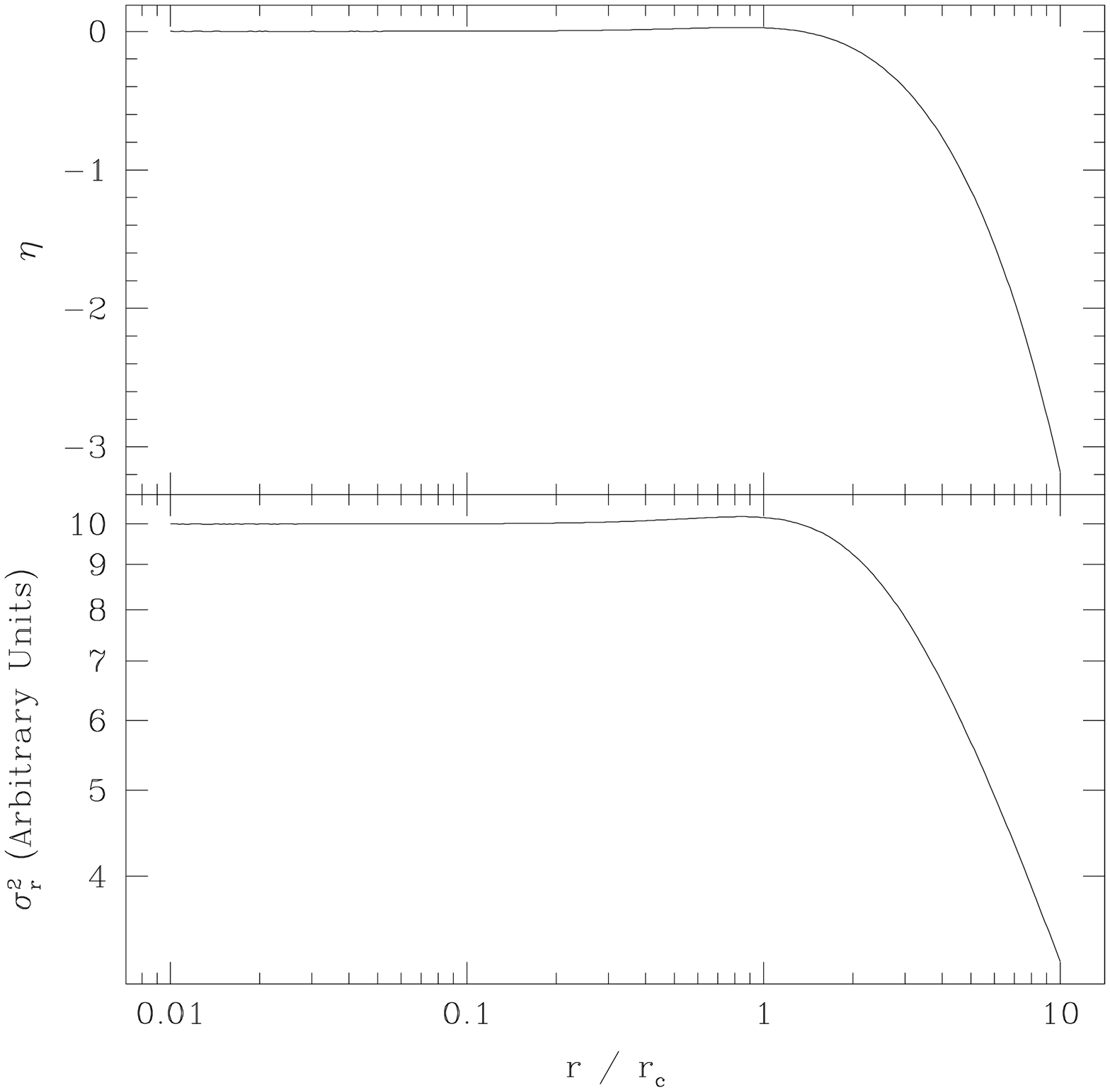}}
\figcaption[isobeta.eps]{Anisotropy profile (top) and radial velocity
dispersion profile (bottom) for the isothermal
\betamodel. \label{fig:isobeta}}

\subsection{Isothermal $\beta$-models}

The mass and emissivity profiles of real clusters are better described
by broken power laws. The \betamodel\ emissivity, which has $\epsilon
\propto (1+x^2)^{-3 \bfit}$, where $x \equiv r / r_c$, fits many
clusters without cooling flows, and is commonly taken to describe an
isothermal gas. If the temperature is constant, and the plasma is in
hydrostatic equilibrium with the gravitational potential, then
according to equation (\ref{eq:iso}) the mass profile has to be
\begin{equation}
M(r) = \frac{3 \bfit k T r_c}{m_p \mu G} \frac{x^3}{1 + x^2},
\label{eq:truebetamass}
\end{equation}
different from the mass profile of the King sphere (Table
\ref{tbl:potentials}). The reason for this discrepancy is that the
classical derivation of the \betamodel\ emissivity assumes that the
King sphere is isothermal (see the Introduction). But the King sphere
is only an approximation to the true nonsingular isothermal sphere
(Binney \& Tremaine 1987, p. 228).  For a \betamodel\ gas to be truly
isothermal, it must be embedded in the mass profile given by equation
(\ref{eq:truebetamass}).

It is then straightforward to derive the orbital properties of the
spherically symmetric dark matter halo using the equations derived in
\S \ref{sec:derive}. They yield $\btrue = \bfit$, and
\begin{equation}
\sigma_r^2(r) = \frac{3 \bfit k T}{4 m_p \mu x^2 } 
\left[ \frac{3 (1 + x^2)^2 \tan^{-1} x}{x (3 + x^2)} - 1 \right].
\end{equation}
The central velocity dispersion is $\sigma_r^2(0) = \beta k T / (m_p
\mu)$, and as $ r \rightarrow \infty$, $\sigma_r^2 \sim 1/r$. Because
the radial velocity dispersion eventually vanishes, the orbits at $r =
\infty$ must be completely tangential to keep the total velocity
dispersion constant, and therefore $\eta \rightarrow -\infty$. The
required anisotropy parameter is independent of $\beta$:
\begin{equation}
\eta(r) = \frac{3}{2} - \frac{2 x^3 (3 + x^2)}
{3 (1+x^2)^2 \tan^{-1} x - x (3 + x^2)}.
\end{equation}
Figure \ref{fig:isobeta} shows the velocity dispersion and 
anisotropy profiles for the isothermal \betamodel.

\subsection{Non-isothermal $\beta$-models}
\label{sec:betamodel}

Although many clusters of galaxies contain an isothermal plasma, a
perhaps larger fraction exhibit temperature gradients. For example,
all but 3 of the 24 clusters examined with \ASCA\ by Markevitch \etal
(1998) have projected temperature profiles which decline with radius.

A temperature gradient results whenever a \betamodel\ emissivity is in
equilibrium with a mass distribution different from equation
(\ref{eq:truebetamass}). Here I calculate the anisotropy solutions for
the \betamodel\ together with the various profiles listed in Table
\ref{tbl:potentials}.

\begin{deluxetable}{ccc}
\tablehead{\colhead{Profile} & \colhead{$\rho(x)$} & \colhead{$M(x)$}
}
\tablecaption{Mass Profiles Considered}
\startdata
SIS           & $ x^{-2}$	      &$x$  
\\
King          & $(1+x^2)^{-3/2}$      &$\ln{(x+\sqrt{1+x^2})}-x/\sqrt{1+x^2}$
\\	      
NFW	      & $x^{-1}(1+x)^{-2}$    &$\ln{(1+x)} - x/(1+x)$  
\\	      
Hernquist     & $x^{-1}(1+x)^{-3}$    &$x^2 (1+x)^{-2} $ 
\\	      
Plummer       & $(1+x^2)^{-5/2}$      &$x^{3} (1+x^2)^{-3/2}$ 

\enddata
\tablecomments{The density and mass profiles are given without
normalization and in terms of
the scaled radius $x \equiv r/r_c$. 
}
\label{tbl:potentials}
\end{deluxetable}

Figure \ref{fig:tprofile} shows the equilibrium temperature profiles
given by equation (\ref{eq:tofr}) for $\bfit = 0.65$ and $\bfit =
1$. Here the temperature always rises towards the core, but only
because $r_c$ is the same for both the emissivity and the matter
density.  Were $r_c$ to be different for the two profiles (as in \S
\ref{sec:cl} below), the temperature could still fall towards the
center. Note that the King approximation to the isothermal sphere,
from which the \betamodel\ is derived, has constant temperature only
within several core radii; at larger radii the plasma temperature
drops steeply.
Conversely, the dark matter potential of a singular isothermal
sphere causes the \betamodel\ gas to have a temperature gradient
inside the core, but a constant temperature at large radii.  As one
would expect, the gas in a cluster with $\bfit = 0.65$ is on the
average hotter than the gas in a cluster of the same mass but with
$\bfit = 1$.

The next step is to decide which values of $\btrue$, the ratio of the
total dark matter energy to the gas energy, are appropriate.  Figure
\ref{fig:badsis} shows the upper and lower limits for the singular
isothermal sphere from equations (\ref{eq:bone}) and
(\ref{eq:btwo}). Because no single value of $\btrue$ can
simultaneously satisfy the limits at all radii, the SIS and the
\betamodel\ are an incompatible potential-emissivity pair.

\vspace{0.1in}
\resizebox{3.5in}{!}{\includegraphics{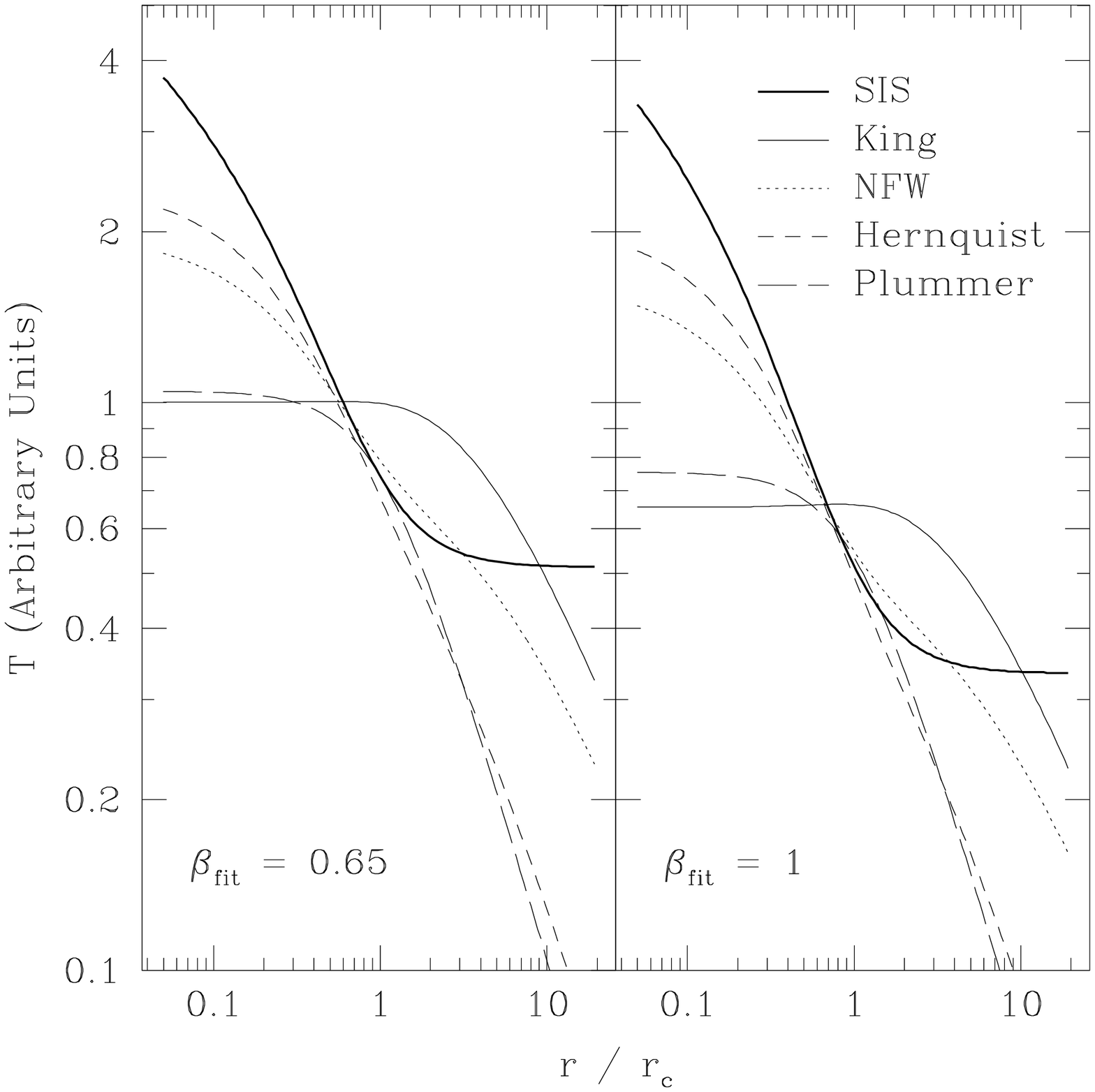}}
\figcaption[tprofile.eps]{The equilibrium temperature profile for two
\betamodel\ emissivity profiles and various gravitational
potentials. The mass profiles are normalized so that they all contain
the same amount of mass within $r = r_c$.
\label{fig:tprofile}}
\vspace{0.1in}

\begin{figure*}
\resizebox{7in}{!}{\includegraphics{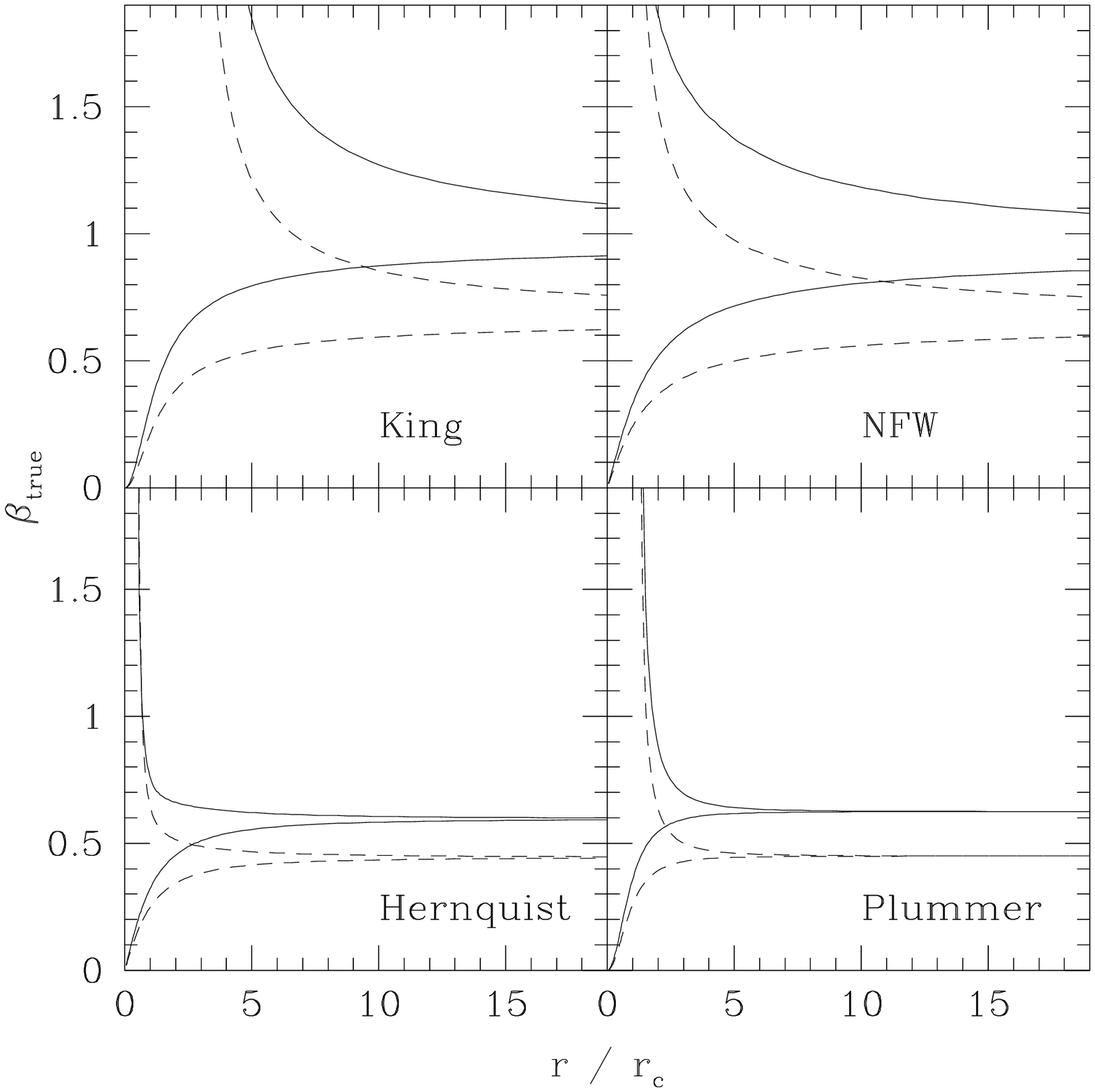}}
\figcaption[betalims.eps]{Upper and lower limits for $\btrue$, the
ratio of the dark matter energy to the gas energy, for a \betamodel\
emissivity combined with the various mass profiles listed in Table
\protect\ref{tbl:potentials}. Solid lines show $\bfit = 1$,
and dashed lines show $\bfit = 0.65$. \label{fig:betalims}}
\end{figure*}

However, as Figure \ref{fig:betalims} shows, the rest of the mass
profiles in Table (\ref{tbl:potentials}) have admissible solutions
when combined with a \betamodel\ gas. In contrast to the scale-free
solutions in \S \ref{sec:powerlaw}, equations (\ref{eq:bone}) and
(\ref{eq:btwo}) allow one unique value of $\btrue$ for each 
potential:
\begin{equation}
\btrue = \lim_{r \rightarrow \infty} \frac{\sigma_2^2}{3 \sigma_1^2}.
\end{equation}

In general, the function $\btrue(\bfit)$, which matches the observed
emissivity, characterized by $\bfit$, with the physical energy ratio
$\btrue$, is not analytic. However, the integrals in equation
(\ref{eq:sigone}) and (\ref{eq:sigtwo}) do have analytic
representations for a \betamodel\ gas embedded in a Plummer sphere.
The exact relations are
\begin{eqnarray}
T(x) & \propto & \frac{4}{3 (4 \bfit + 1) \sqrt{1 + x^2}}, \\
\btrue &  = & \frac{4 \bfit + 1}{8}, \\
\sigma_r^2(x) & \propto & \frac{1}{6 \sqrt{1 + x^2}} \\
\eta(r) & = & 0,
\end{eqnarray}
where $x \equiv r / r_c$. 

The Plummer model is remarkably well matched to the \betamodel\ gas
emissivity. Its velocity dispersion is always independent of $\bfit$,
and the dark matter orbits are completely isotropic. This result is
surprising and counterintuitive, because the \betamodel\ emissivity is
historically derived from the isotropic King model.  But as I have
discussed above, that derivation assumes that the perfectly isotropic
King sphere is isothermal, an assumption which breaks down well
outside the core. In contrast, though it is not isothermal, a
\betamodel\ gas embedded in an $\eta = 0$ Plummer sphere satisfies the
equations of hydrostatic equilibrium perfectly.

\begin{figure*}
\resizebox{7in}{!}{\includegraphics{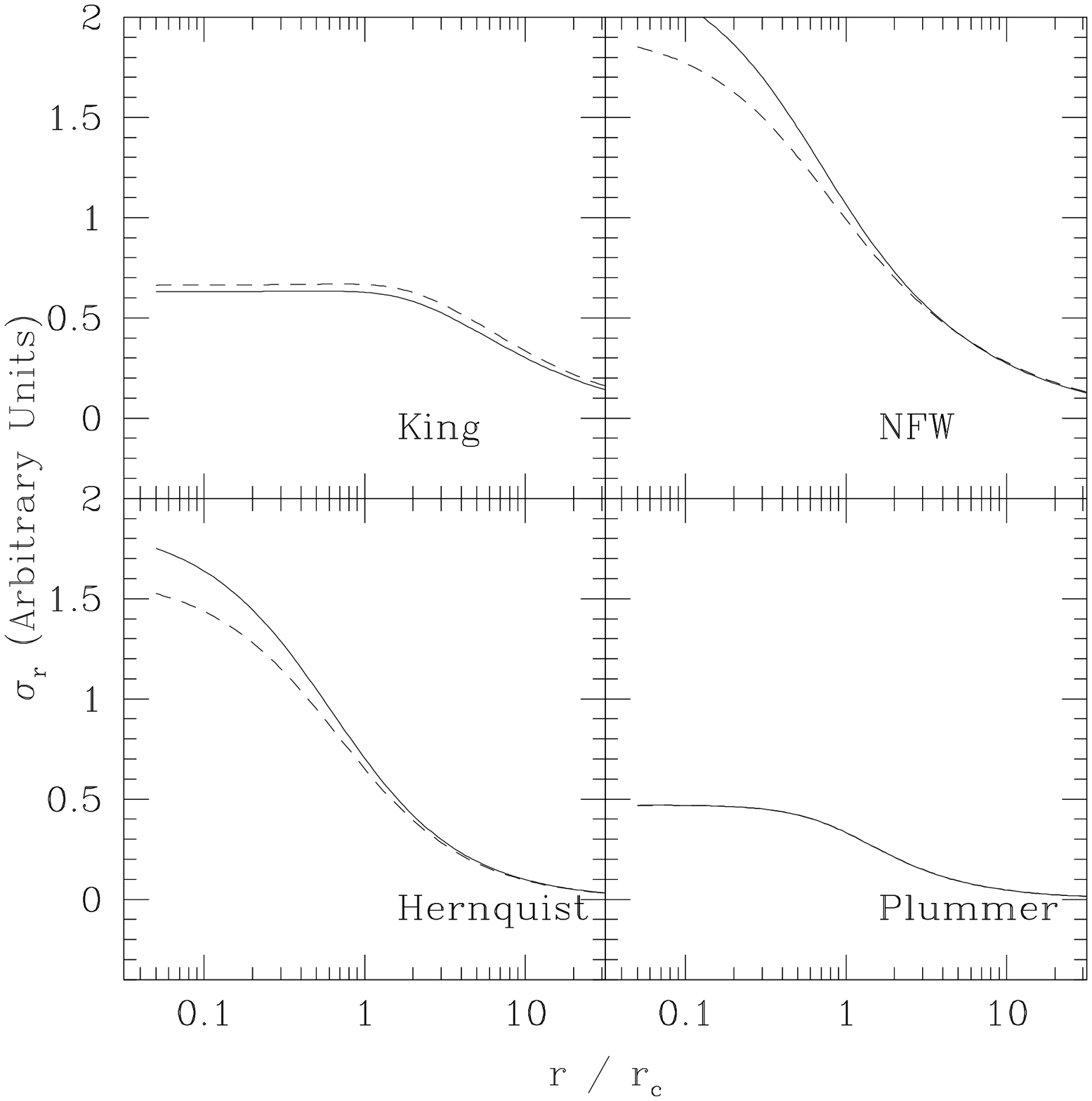}}
\figcaption[betasig.eps]{Radial velocity dispersion profiles for the
various potentials in Table \protect\ref{tbl:potentials} when they
contain a \betamodel\ gas. Solid lines show $\bfit = 1$, and dashed
lines show $\bfit = 0.65$. \label{fig:betasig}}
\end{figure*}

\begin{figure*}
\resizebox{7in}{!}{\includegraphics{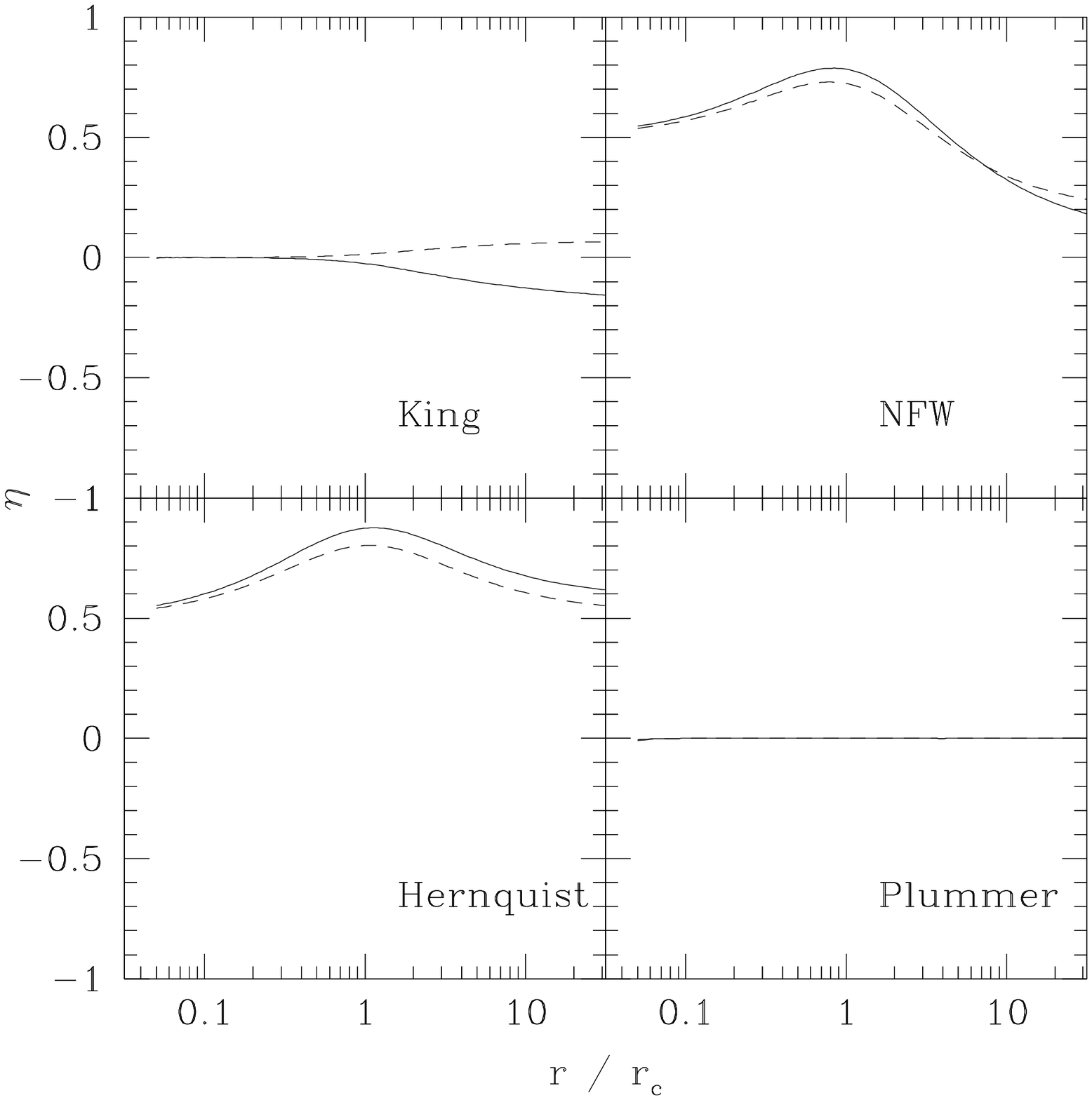}}
\figcaption[betaeta.eps]{Anisotropy profiles for the various
potentials in Table \protect\ref{tbl:potentials} when they contain a
\betamodel\ gas. Solid lines show $\bfit = 1$, and dashed lines show
$\bfit = 0.65$. \label{fig:betaeta}}
\end{figure*}

For the King, NFW, and Hernquist models $\btrue(\bfit)$ is also
roughly linear, with the general property that $\btrue < \bfit$ in
each case. The resulting radial velocity dispersion profiles, given by
equation (\ref{eq:sigma}), and anisotropy profiles, given by equation
(\ref{eq:anisotropy}), appear in Figures \ref{fig:betasig} and
\ref{fig:betaeta}, respectively. For the King profiles, $\beta$
correlates strongly with the orbital anisotropy at infinity; $\bfit =
1$ gives almost completely radial orbits at infinity, $\bfit = 0.65$
results in orbits with a greater tangential component, and $\beta =
3/4$ gives perfect isotropy at all radii.

\vspace{0.2in}
\resizebox{3in}{!}{\includegraphics{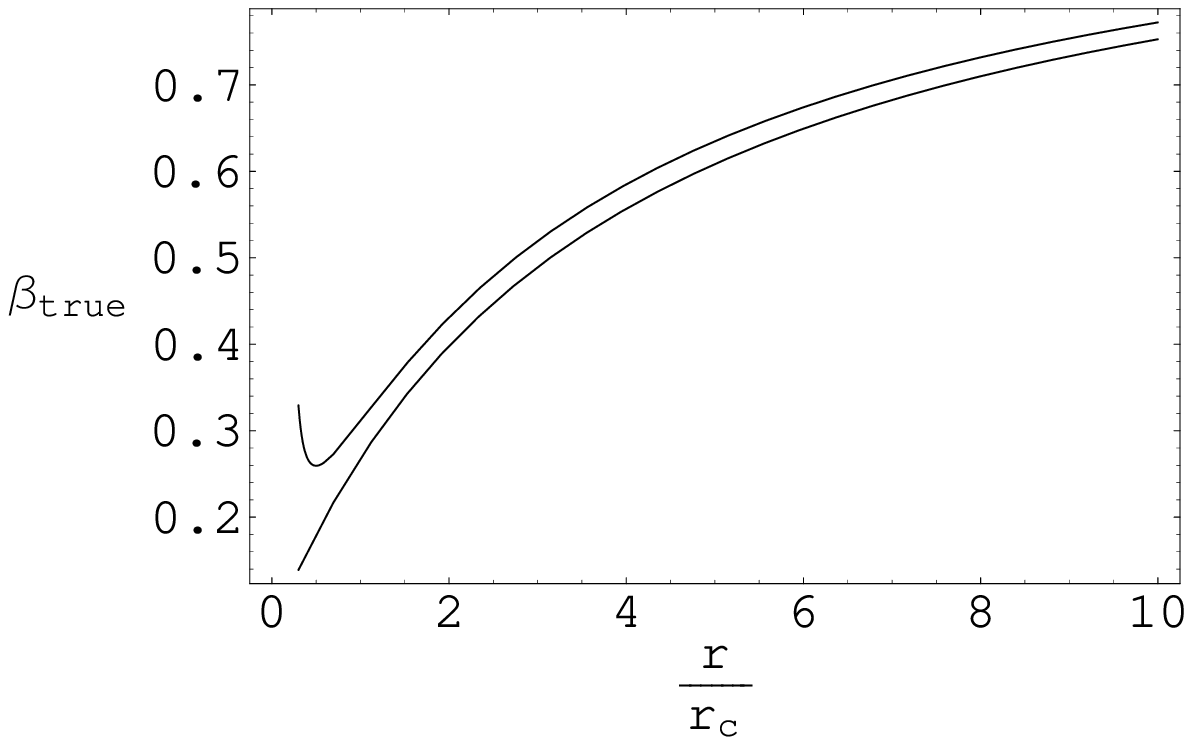}}
\figcaption[badsis.eps]{Upper and lower limits for $\btrue$,
the ratio of the dark matter energy to the gas energy, for
a \betamodel\ gas embedded in a singular isothermal sphere.
The figure shows that a constant $\btrue$ is not allowed.
 \label{fig:badsis}}
\vspace{0.2in}

The anisotropy profiles of the NFW and Hernquist models show a
particularly interesting structure. Both profiles have a small radial
anisotropy at $r = 0$, attain a peak of $\eta \approx 0.75$ near $r =
r_c$, and then decline towards a final, positive value.

The King and Plummer spheres exhibit featureless $\eta(r)$, whereas
the NFW and Hernquist anisotropies have a maximum value. The physics
behind this difference lies in the behavior of the dark matter density
as $r \rightarrow 0$. The NFW and Hernquist profiles have a cuspy
$\rhodm \propto r^{-1}$ for $r \ll r_c$, whereas the King and Plummer
spheres have finite density at $r = 0$. If the gas is in perfect
hydrostatic equilibrium with the cuspy models, its temperature profile
rises as $r \rightarrow 0$. For the dark matter to gas energy ratio
$\btrue$ to be constant, the total velocity dispersion
$\sigma_\mr{tot}^2 = \sigma_r^2 ( 3 - 2 \eta)$ must then rise towards
the center as well. But $\sigma_r^2$ is also coupled to the
temperature and mass distribution via the Jeans equation
(\ref{eq:jeans}). For $r < r_c$, the temperature term $\sigma_1^2$
from equation (\ref{eq:sigone}) dominates. Because of the density cusp
at $ r < r_c$, $\sigma_1^2$ itself falls more slowly than the
temperature, and an increasing radial anisotropy is required to
compensate, keeping $\btrue$ constant. At larger radii $r > r_c$,
however, the subtractive, gravitational term $\sigma_2^2$ competes
with the temperature term and causes $\sigma_r^2$ to fall rapidly. A
larger tangential velocity component is then required to maintain a
constant dark matter to gas energy ratio, and $\eta(r)$ falls.

Remarkably, the NFW and Hernquist anisotropies in Figure
\ref{fig:betaeta} are similar to the dark matter anisotropy profiles
in N-body simulations without gas (Cole \& Lacey 1996; Kauffman \etal
1999; Diaferio 1999).  In these simulations, $\eta(r)$ is positive at
$r = 0$, achieves a maximum near the virial radius, and then decreases
to negative values. The close correspondence between the results of
this derivation, which rely closely on the structure of the X-ray
emitting gas in the dark matter halos, and the N-body simulations,
which do not include gas physics, is encouraging.

\subsection{Application to CL0024+16}
\label{sec:cl}

Now I use X-ray imaging, gravitational lensing, and an optical
redshift survey to test the simple hydrostatic formalism developed
above. I derive unique velocity dispersion and anisotropy profiles by
combining X-ray and lensing data for CL0024+16. I calculate the shape
and normalization of the dark matter velocity dispersion profile
without recourse to any free parameters. This profile, once projected,
can be directly compared to the independent galaxy velocity dispersion
profile, providing a powerful consistency test of the complete
dynamical model.

CL0024+16, a $z = 0.39$ cluster, has \ROSAT\ High Resolution Imager
X-ray data (B\"ohringer \etal 2000), an \ASCA\ temperature measurement
(Soucail \etal 2000), and a mass profile from strong lensing (Tyson,
Kochanski, \& Dell'Antonio 1998). These observations independently fix
the X-ray emissivity and the mass profile.  Furthermore, a catalog of
optical redshifts for 138 galaxies in the field of the cluster is
available (Dressler \etal 1999).

The Tyson \etal (1998) surface mass density is characterized by
a soft core within the lensing radius $r_l$:
\begin{equation}
\Sigma(R)  =  K_1 \frac{(1 + \gamma R^2/r_l^2)}{(1 +
R^2/r_l^2)^{2-\gamma}},
\label{eq:mascon}
\end{equation}
where $R$ is the projected distance from the cluster center, $K_1 =
7900\hhh M_\odot$ pc$^{-2}$ is the central surface density, the slope 
$\gamma = 0.57$, and $r_l = 35\hhh$\m\ kpc, with the Hubble Constant
$H_0 = 100 \hhh$ km s\m\ Mpc\m. Outside the lensing radius $r_l$
the mass profile resembles an NFW profile with a core radius $r_c 
= 311\hhh$\m\ kpc:
\begin{equation}
\Sigma(R) =  \frac{M}{2 \pi r_c^2 (R^2/r_c^2 - 1)} \left(1 - 
\frac{\mr{sec}^{-1} R/r_c}{\sqrt{R^2/r_c^2 - 1}}\right).
\end{equation}
Here $M$ is determined by requiring continuity with equation
(\ref{eq:mascon}). Thus the density profile of CL0024+16 resembles
that of a NFW halo, but lacks a central singularity.

The deprojected density profile is given by (Binney \& Tremaine 1987, 
p. 205):
\begin{equation}
\rhodm(r) = -\frac{1}{\pi}
 \int_r^\infty \dd{\Sigma}{R} \frac{dR}{\sqrt{R^2 - r^2}}.
\end{equation}
This integral is well approximated by the function,
\begin{equation}
\rhodm(r) = \frac{107.5 \hhh^2}{\left(
1+ r/r_0 \right)^{3}}
\times 10^{15}
M_\odot \mr{\ Mpc}^{-3},
\end{equation}
where $r_0 = 0.0725 \hhh$\m\ Mpc. The resultant projected mass profile
is better than 12\% accurate everywhere within the lensing arcs,
commensurate with the $\approx 15\%$ error in the Tyson \etal (1998)
parameters. The X-ray emissivity, on the other hand, is well described
by a \betamodel\ with core radius $33 \hhh$\m\ kpc and $\bfit = 0.475$
(B\"ohringer \etal 2000).

\begin{figure*}
\resizebox{7in}{!}{\includegraphics{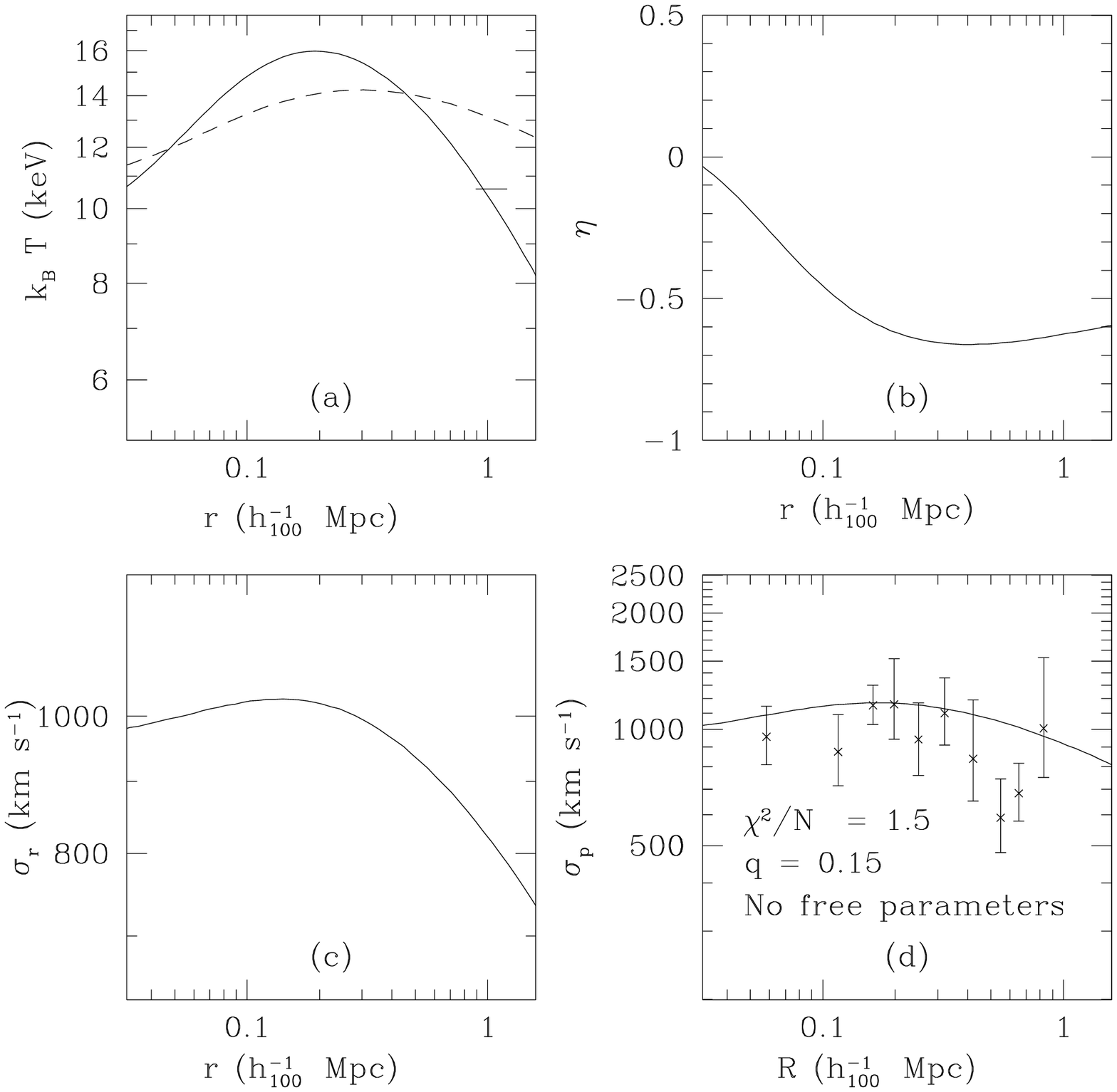}}
\figcaption[cl0024.eps]{Derived properties of CL0024+16, and a
comparison with the optical spectroscopic data: (a) predicted plasma
temperature (solid line) and mean integrated emission-weighted
temperature as observable by \ASCA\ (dashed line); (b) predicted
orbital anisotropy profile; (c) predicted three-dimensional radial
velocity dispersion; (d) predicted line-of-sight velocity dispersion
profile along with the measured galaxy velocity dispersion (Dressler
\etal 1999).
\label{fig:cl0024}}
\end{figure*}

Because the central cooling time of the CL0024+16 is approximately the
same as the age of the cluster, it should not contain a cooling flow
(Soucail \etal 2000). Therefore the hydrostatic analysis discussed in
the previous sections is appropriate. I take the mean molecular weight
to be $\mu = 0.6$. Note that of all the properties derived by the
model, only the temperature depends on the choice of $\mu$. Figure
\ref{fig:cl0024} shows the results.

According to the model, the central temperature of the cluster should
be $\approx 9.2$ keV. The emission-weighted temperature within the
\ASCA\ aperture should be 12 keV; the Soucail \etal (2000)
spectroscopic analysis gives a 90\% confidence interval of 3.6--10.6
keV. Thus the predicted plasma temperature is consistent with but
somewhat above the measurement. The discrepancy is not tied to the
present analysis, but is an intrinsic property of the lensing and the
X-ray data, which give only marginally consistent mass measurements
for the cluster (Soucail \etal 2000). The lensing mass measurement
might be an overestimate because of contamination by large scale
structure projected along the line of sight (Metzler \etal 1999). The
X-ray measurement is also suspect, because $\beta < 0.5$ in theory
yields an infinite X-ray luminosity for the cluster.

Given the X-ray and lensing data, however, the calculations show that
the dark matter to gas energy ratio is $\btrue = 0.561$. Then
it is straightforward to derive the anisotropy profile $\eta(r)$ and
the radial velocity dispersion profile $\sigma_r^2(r)$ through
equations (\ref{eq:sigone})--(\ref{eq:sigtwo}). The line-of-sight dark
matter velocity dispersion profile is (Binney \& Tremaine
1987, p. 208)
\begin{equation}
\sigma_p^2(r) = \frac{2}{\Sigma} \int_R^\infty \left(1 - \eta
\frac{R^2}{r^2} \right) \frac{\nu \sigma_r^2 r dr}{\sqrt{r^2 - R^2}}.
\label{eq:project}
\end{equation}
The theoretical $\sigma_p(r)$ appears in Figure \ref{fig:cl0024}
alongside the line-of-sight velocity dispersion profile of the
galaxies. To calculate $\sigma_p(r)$ from the Dressler \etal (1999)
data, it is first necessary to determine the well-mixed cluster
membership. I first select all Dressler \etal (1999) galaxies in the
CL0024+16 field with redshifts $0.381 < z < 0.404$, corresponding to a
$\pm 2500$ km s\m\ line-of-sight velocity relative to the cluster rest
frame. Then I apply the DEDICA algorithm (Pisani 1993) to determine
the cluster membership. DEDICA uses Gaussian kernels to arrive at a
maximum-likelihood estimate of cluster membership, in this case 104
galaxies. Finally, I measure $\sigma_p(r)$ by gathering the members
into 10 bins of 10-11 galaxies,
and calculating $\sigma_p$ and its associated errors using standard
bootstrap analysis.

The simple model discussed above well describes the velocity
dispersion profile of CL0024+16. The $\chi^2$ statistic is 15 for 10
degrees of freedom, yielding an acceptable fit quality $q = 0.15$
(Press \etal 1993, p. 660). The agreement is remarkable given that no
free parameters are allowed in the comparison.

\section{Multiphase Cooling Flow Solutions}
\label{sec:cooling}

\subsection{General Properties}

\resizebox{3.5in}{!}{\includegraphics{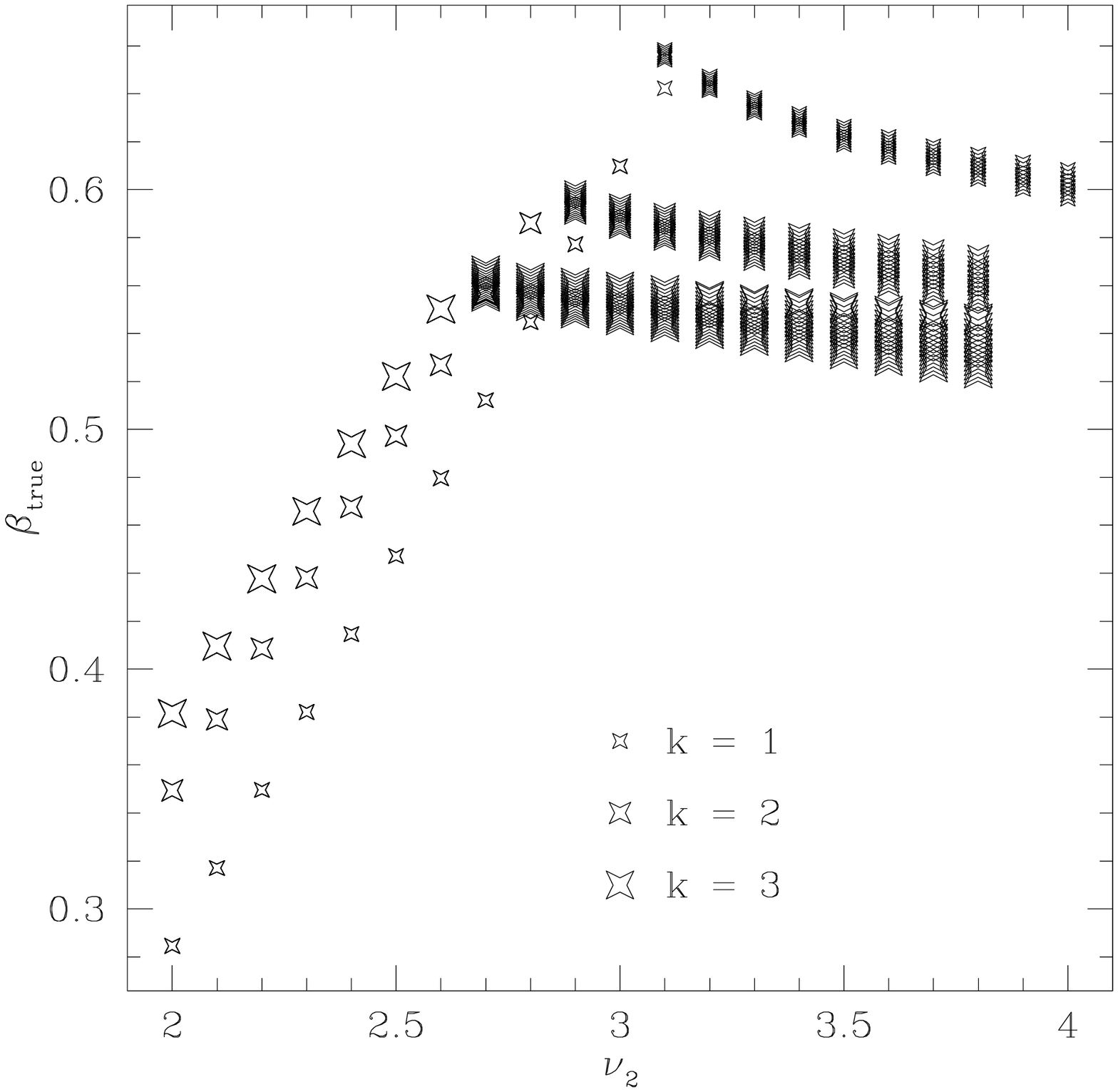}}
\figcaption[coolbtrue.eps]{The dark matter to gas energy ratio,
$\btrue$, for cooling flow models with $k =$ 1, 2, 3 and faint-end
emissivity slopes $\nu_2$.
\label{fig:coolbtrue}}
\vspace{0.2in}

So far this discussion has focused on a intracluster plasma in
hydrostatic equilibrium with the gravitational potential. This
scenario provides a good description of most clusters, at least for
gas removed more than $\approx 100$-200 kpc from the cluster
center. However, the plasma in many clusters has a cooling time
significantly shorter than the Hubble time within $r \approx 100$ kpc,
an indication that an inward flow may be present. It is worth
exploring how the dark matter anisotropy changes within the cooling
regime.

Here I apply the formalism derived in \S \ref{sec:derive} to clusters
which contain a cooling flow. In this case, hydrostatic equilibrium no
longer applies, and equation (\ref{eq:hydro}) must be replaced with a
set of cooling flow equations. I adopt the multiphase cooling flow
equations discussed by Thomas (1998), who examines a quasihydrostatic,
spherically symmetric flow with an emulsion of comoving but thermally
isolated density phases. The Thomas (1998) scenario, summarized
briefly in the Appendix, is a refinement of the classical multiphase
models of Nulsen (1986).

\begin{figure*}
\resizebox{7in}{!}{\includegraphics{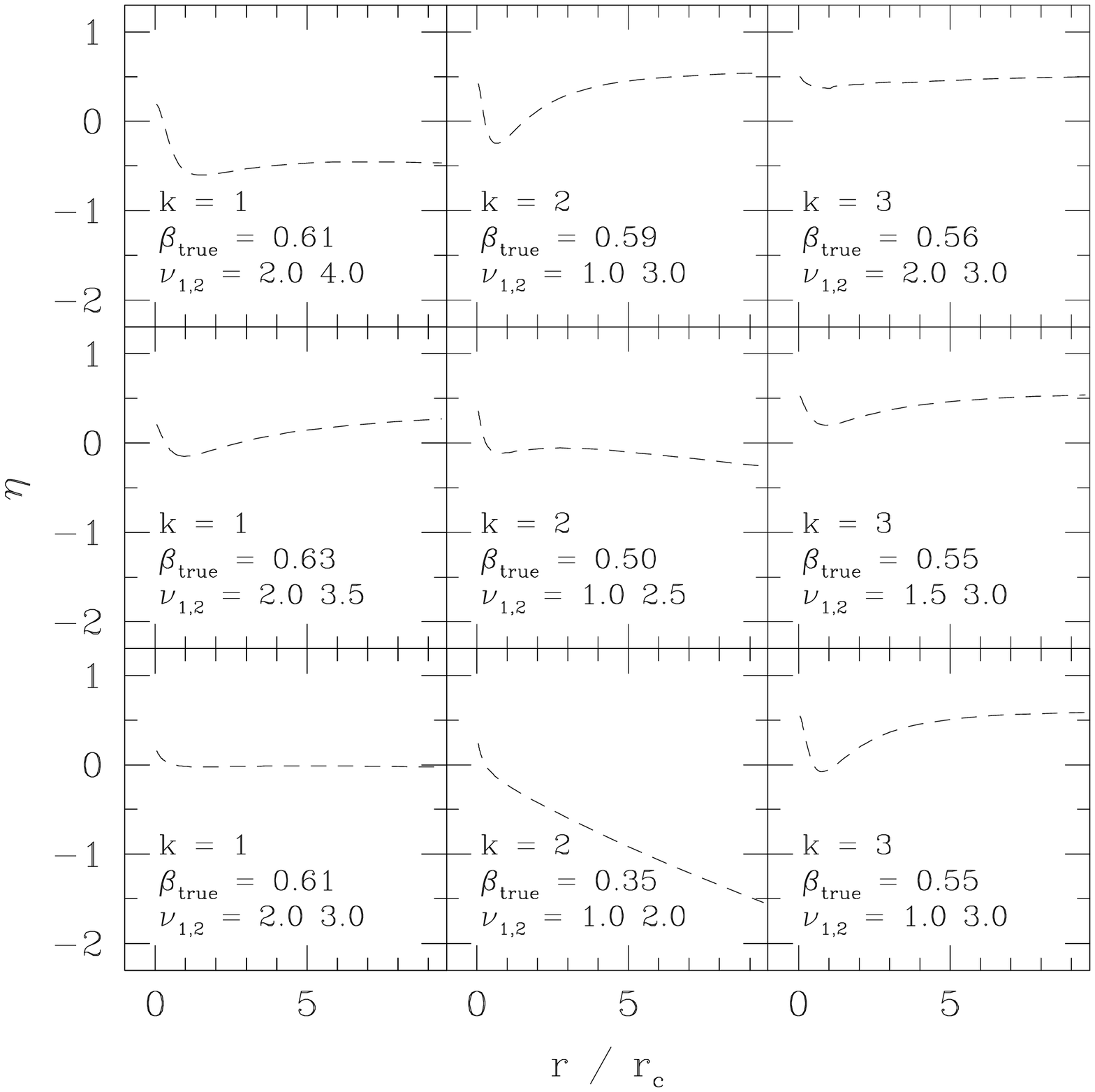}}
\figcaption[cooleta.eps]{Anisotropy profiles for dark matter halos containing
quasihydrostatic cooling flows with the indicated $k$-parameters.  The
emissivity profile characterized by $\nu_1$ and $\nu_2$ is given
in equation (\protect\ref{eq:coolem}). Also shown is
$\btrue(k,\nu_1, \nu_2)$. \label{fig:cooleta}}
\end{figure*}

The cooling flow equations have a free nondimensional parameter, $k$,
which together with the gas emissivity $\epsilon$ completely specifies
the shapes of the flow temperature, the mass deposition rate, and the
gravitational potential. In the Thomas (1998) model, $k$ describes, in
an average sense, the nature of the phase mixture in the flow, and can
have values from 1 to $\infty$. Emulsions with $k = \infty$ include
phases of arbitrarily low density; those with $k \sim 1$ allow only a
small range of densities at each radius. Thomas (1998) argues that the
$k \sim 1$ models are preferable because they reproduce observations
of many cooling flow clusters where the emission-weighted temperature
drops as $r \rightarrow 0$; on the other hand, the $k = \infty$
solutions have increasing $T(r)$ as $r \rightarrow 0$, in contrast
with the observations.

Once $k$ and the emissivity $\epsilon$ are specified, the resulting
temperature and mass profiles constrain the anisotropy profile of the
dark matter halo through equations (\ref{eq:sigma})--(\ref{eq:eta}).
I will not use the \betamodel\ emissivity for the cooling flow
calculations, first because it does not accurately match the observed
emissivity profiles of cooling flow clusters, which are often cuspy as
$r \rightarrow 0$, and also because it is inconsistent with all but
the $k = \infty$ flows. Instead I use the broken power law
\begin{equation}
\epsilon(x) \propto \left( x^{\nu_1} + x^{\nu_2} \right)^{-1},
\label{eq:coolem}
\end{equation}
where $x \equiv r / r_c$ and $\nu_2 > \nu_1$. An emissivity
profile with $\nu_1 = 0$ resembles a traditional \betamodel\ profile
with $\bfit = \nu_2/6$.  See the Appendix for a detailed calculation.

Again, the first step in deriving the anisotropy profile is
calculating $\btrue$, the constant dark matter to gas energy ratio.
Figure \ref{fig:coolbtrue} shows the results.  Here $\btrue$ depends
strongly on the faint-end slope of the emissivity, $\nu_2$. When
$\nu_2 \simless 3$, $\btrue$ is completely independent of $\nu_1$; for
$\nu_2 \simgreat 3$, a slight perturbation exists as a function of
$\nu_1$.



One interesting property of the solutions is that $\btrue$ never
exceeds 0.65, because the multiphase equations do not allow larger
values. The $k \sim 1$ models, which best match real clusters (Thomas
1998), allow the largest range of $\btrue$; the $k = \infty$ models
always have $0.4 < \btrue < 0.5$.

Figure \ref{fig:cooleta} shows the resulting anisotropy profiles.  A
wide variety of profiles occur depending on the specific value
of $\nu_1$, $\nu_2$, and $k$. However, all $\eta(r)$ begin positive
and decline. When the emissivity is shallow---$\nu_2 = 2$, for
example---the anisotropy tends towards negative values at larger
radii. However, for reasonably steep emissivity profiles, the
anisotropies stay constant or increase at large radii, and exhibit a
substantial radial bias outside the core.

Although the cooling flow solutions extend to $r = \infty$, they have
the greatest validity inside the cooling radius. If the measured
emissivity of a cluster drops off very steeply at large radii ($\nu_2
\simgreat 4$), no multiphase cooling flow solution can properly
account for the gas physics at all radii. In such cases it is best to
regard the true anisotropy parameter as resembling the cooling flow
solution within the core, and the hydrostatic solutions (\S
\ref{sec:betamodel}) for $r > r_c$.

\vspace{0.2in} \resizebox{3.5in}{!}{\includegraphics{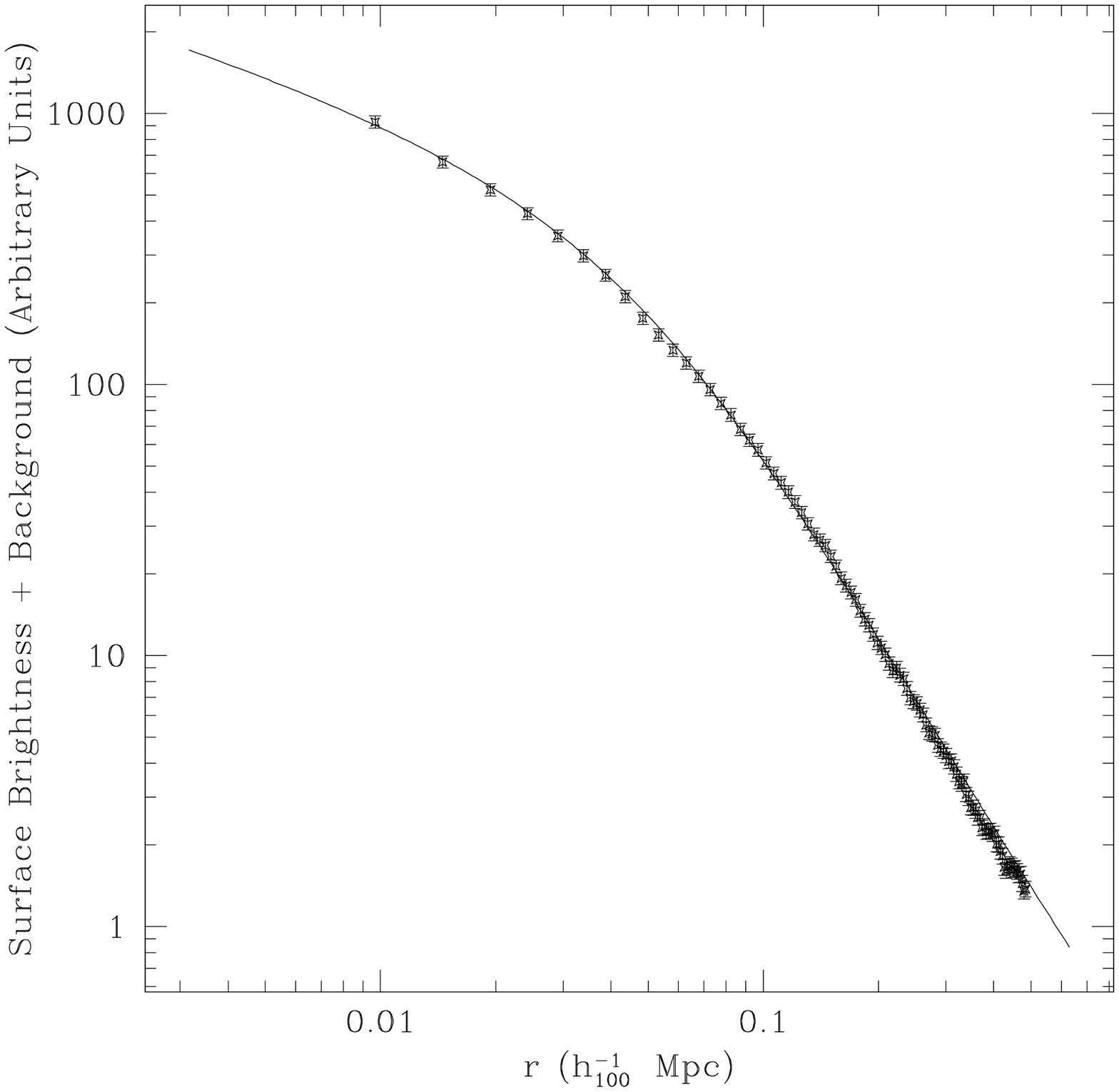}}
\figcaption[a2199em.eps]{Surface brightness plus background for the
\ROSAT\ PSPC observation of Abell 2199. The error bars show the
uncertainty due to photon noise plus a 5\% additional error to account
for uncertainty in the location of the bins. The solid line shown is
the best fit combination of a constant background plus the projection
of the broken power law from equation (\protect\ref{eq:coolem}) along
the line of sight, with $\nu_1 = 1.35$, $\nu_2 = 3.41$, and $r_c =
0.059\hhh$\m\ Mpc. \label{fig:a2199em}}
\vspace{0.2in}

\subsection{Application to Abell 2199}

Here I test the cooling flow anisotropy model by applying it to X-ray
and optical observations of Abell 2199. I predict the shape and
normalization of the projected velocity dispersion profile from the
X-ray data, and compare it directly with independent optical
spectroscopy. Here the cooling flow model determines the shapes of the
mass and temperature profiles from the X-ray emissivity, and no
additional lensing mass is required. An absolute temperature
measurement provides the normalization.

Abell 2199 is a $z = 0.03$ cluster that contains a moderate cooling
flow, with a mass deposition rate $\dot{M} \approx 150 M_\odot$ yr\m\
(Peres \etal 1998). The \emph{Einstein} observatory emission-weighted
temperature is $4.5 \pm 0.2$ keV within $2\hhh$\m\ Mpc (David \etal
1993). Velocity measurements for 98 galaxies in the field of the
cluster are also available (Hill \& Oegerle 1998).

To determine the shape of the X-ray emissivity, I examine a publicly
available \ROSAT\ PSPC observation of Abell 2199, with sequence
identification RP800644N00, and a 41000 s exposure time. I use the
MIDAS/EXSAS software package (Zimmermann \etal 1993) to calculate the
surface brightness profile within $0.5 \hhh$\m\ Mpc. I then fit a
constant background plus the surface brightness profile given by the
projection of equation (\ref{eq:coolem}) along the line of sight. A
deprojected emissivity with $\nu_1 = 1.35$, $\nu_2 = 3.41$, and $r_c =
0.06\hhh^{-1}$ Mpc provides an excellent fit with a reduced chi squared
equal to 0.6 (Figure \ref{fig:a2199em}). The errors on the fitted
parameters are less than 10\%. Note that the innermost bin (not shown)
is omitted from the analysis to minimize smearing of the emission due
to the PSPC point spread function.

\begin{figure*}
\resizebox{7in}{!}{\includegraphics{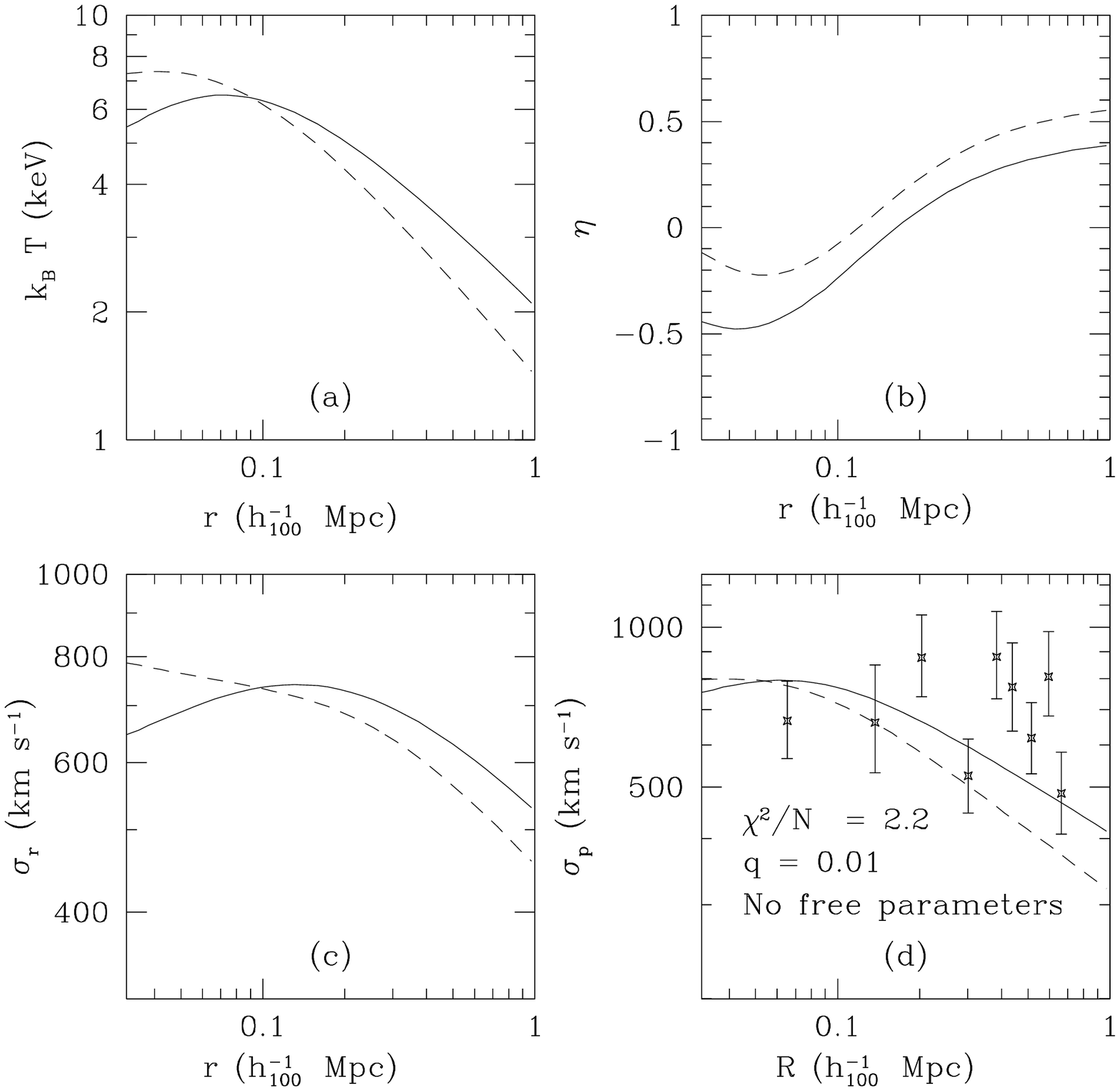}}
\figcaption[a2199.eps]{Derived properties of Abell 2199 for cooling
flow solutions with $k = 1$ (solid lines) and $k = 2$ (dashed lines):
(a) predicted plasma temperature; (b) predicted orbital anisotropy
profile; (c) predicted three-dimensional radial velocity dispersion;
(d) predicted line-of-sight velocity dispersion profile along with the
measured galaxy velocity dispersion (Hill \& Oegerle 1998).
\label{fig:a2199}}
\end{figure*}

Once $\nu_1$ and $\nu_2$ are specified, the shapes of the temperature
and mass profiles are fixed by the cooling flow equations. The David
\etal (1993) temperature measurement fixes their normalization, and it
is possible to fully calculate the velocity dispersion and anisotropy
profiles. The line-of-sight velocity dispersion profile is then given
by equation (\ref{eq:project}). To measure the profile independently
from the Hill \& Oegerle (1998) data, I apply the same procedure as
that described in \S \ref{sec:cl}. The results of the analysis for
both the $k = 1$ and $k = 2$ solutions are shown in Figure
(\ref{fig:a2199}). For $k = 1$ $\btrue = 0.62$, and for $k = 2$
$\btrue = 0.57$. The profiles are steeper for larger values of $k$
as $r \rightarrow \infty$.

Once again, this simple formalism provides an adequate description of
the measured velocity dispersion profile without recourse to any free
parameters. The reduced chi squared is 2.2, providing an acceptable
fit for 9 degrees of freedom. Solutions with $k > 1$ are excluded by
the data, because the velocity dispersion profile falls too quickly,
affirming the intuition of Thomas (1998) that multiphase cooling flows
with a broad density phase distribution are less likely to be
consistent with real clusters.

\section{Conclusion}
\label{sec:conclusion}
I use the X-ray emission from hot gas embedded in a dark matter halo
to break the degeneracy between the anisotropy parameter $\eta$ and
the system mass.  Given an assumed mass profile and and X-ray
emissivity, the model simultaneously determines the anisotropy profile
and the radial velocity dispersion profile. The resultant $\eta(r)$
for the hydrostatic case are consistent with anisotropy profiles from
N-body simulations without gas. With the addition of central cooling
flows, the formalism predicts a rich variety of anisotropy profiles
within the cooling radius, most of which, however, contain a
substantial radial anisotropy at the cluster center and at large
radii.

I apply the formalism to two sets of observations of clusters of
galaxies: the lensing cluster CL0024+16 and the cooling flow cluster
Abell 2199. In both cases, the velocity dispersion profiles predicted
by the anisotropy models are consistent with independently measured
optical data. The agreement is particularly encouraging because no
free parameters are allowed in the comparisons.

The warning of Binney \& Tremaine (1987, p. 208), that for spherical
systems, ``radically different models can be consistent with both the
Jeans equations and the observations,'' is particularly applicable
here and should not be neglected. This paper shows, however, that a
combination of the X-ray emissivity and other auxiliary data, including
the lensing mass or the gas temperature, provides a useful and
self-consistent starting point for constraining the orbital structure
of dark matter halos with gas.

This research was supported by the Smithsonian Institution. I thank
Margaret Geller, Joseph Mohr, and the anonymous referee for
suggestions that improved the paper considerably.  I am grateful to
Lars Hernquist, Kathleen Kang, and Daniel Koranyi for useful
discussions.

\clearpage
\appendix \section{Derivation of the Anisotropy Profile from the Cooling Flow
Equations} 

Thomas (1998) derives equations that describe a quasihydrostatic,
spherically symmetric flow consisting of an emulsion of comoving but
thermally isolated density phases. The relevant variables are the
mass, $M$, the temperature, $T$, the emissivity, $\epsilon$, the mass
accretion rate, $\mdot$, and the following dimensionless quantities:
\begin{eqnarray}
\Sigma & \equiv & \frac{G M \mu m_p}{2 r k_B T}, \\
\tau   & \equiv & \frac{1}{k} \dlndln{\mdot}{r}, \\
\chi   & \equiv & \dlndln{M}{r},
\end{eqnarray}
where $k_B$ is Boltzmann's constant, and $k$ is a dimensionless
parameter that characterizes the range of densities present at
each radius. The $k = 1$ models possess a minimum density at each
radius and are the least extended. The $k = \infty$ models include
phases of arbitrary low density, and are the most extended
convectively stable distributions.

With the assumption that the cooling function is dominated by thermal
bremsstrahlung ($\Lambda \propto \rho_g^2 T^{1/2}$) and that the ratio
of specific heats $\gamma = 5/3$, the Thomas (1998) steady-state
equations have the solutions
\begin{eqnarray}
\tau   & = & \frac{r^3 \epsilon^{5/7}}{
		\left(20/21 + k \right) \int r^2 \epsilon^{5/7} dr}, \\
\Sigma & = & - \frac{5}{14} \left(\tau + \dlndln{\epsilon}{r} \right), \\
\chi   & = & 1 - \frac{4}{5} \Sigma + \frac{2}{3} \tau + \dlndln{\Sigma}{r}.
\end{eqnarray}
Rearranging the equations to solve for the temperature yields
\begin{equation}
T \propto \epsilon^{2/7} \left[ \left( \frac{20}{21} + k \right)
		\int r^2 \epsilon^{5/7} dr \right]^\frac{20}{20+21k},
\end{equation}
where the expression in square brackets approaches 1 as $k \rightarrow
\infty$. It is therefore evident that for all reasonable emissivity
profiles, the $k = \infty$ solutions have $dT / dr < 0$ everywhere.

Once the temperature is fixed, it is possible to constrain $\btrue$,
the ratio of the total dark matter kinetic energy to the gas energy.
Because $4 \pi r^3 \rhodm = M \chi$, the velocity dispersion equations
(\ref{eq:sigone})--(\ref{eq:sigtwo}) become
\begin{eqnarray}
\label{eq:sigcool1}
\sigma_1^2 &=& \frac{k_B}{m_p \mu r T \Sigma \chi}
		\int 3 \btrue T^2 \Sigma \chi dr,\\
\label{eq:sigcool2}
\sigma_2^2 &=& \frac{k_B}{m_p \mu r T \Sigma \chi} 
		\int 2 T^2 \Sigma^2 \chi dr.
\end{eqnarray}
In cases where $\sigma_1^2 \gg k_B T/(m_p \mu)$ at large radii, $\btrue$
once again takes on a unique value:
\begin{equation}
\btrue = \lim_{r \rightarrow \infty} \frac{\sigma_2^2}{3 \sigma_1^2}.
\end{equation}
If the integrals in equations (\ref{eq:sigcool1})-(\ref{eq:sigcool2})
diverge as $r \rightarrow \infty$, then, applying l'H\^opital's rule,
\begin{eqnarray}
\btrue & = & \frac{2}{3} \lim_{r \rightarrow \infty} \Sigma. \\
       & = & -\frac{5}{21} \left[ \frac{3}{20/21 + k} + 
		\lim_{r \rightarrow \infty} \dlndln{\epsilon}{r}
		\left( \frac{5}{20/3 + 7 k} + 1 \right) \right].
\end{eqnarray}
If, on the other hand, the integrals converge, then $\btrue$ must be
determined numerically. After this, direct application of equations
(\ref{eq:sigma}) and (\ref{eq:eta}) will yield the radial velocity
dispersion profile $\sigma_r$ and the anisotropy profile $\eta(r)$.


\begin{references}
Binney, J., and Tremaine, S. 1987, Galactic Dynamics (Princeton:
Princeton University Press)

B\"ohringer, H., Soucail, H., Mellier, Y., Ikebe, Y., \&
Schuecker, P. 2000, A\&A, 353, L24

Carlberg, R. G., Yee, H. K. C., Ellingson, E., Morris, S. L., Abraham, 
R., Gravel, P., Pritchet, C. J., Smecker-Hane, T., Hartwick, F. D. A., 
Hessler, J. E., Hutchings, J. B., \& Oke, J. B. 1997, ApJ, 495, L13

Cole, S., \& Lacey, C. 1996, MNRAS, 281, 716

David, L. P., Slyz, A., Jones, C., Forman, W., Vrtilek, S. D., \&
Arnaud, K. A. 1993, ApJ, 412, 479

Diaferio, A. 1999, MNRAS, 309, 610

Dressler, A., Smail, I., Poggianti, B. M., Butcher, H., Couch, W. J.,
Ellis, R. S., and Oemler, A. , Jr. 1999, ApJS, 122, 51

Eke, V., Navarro, J. F., \& Frenk, C. S. 1998, ApJ, 503, 569

Geller, M. J., Diaferio, A., \& Kurtz, M. J. 1999, ApJ, 517L, 23

Gerhard, O., Jeske, G., Saglia, R. P., \& Bender, R. 1998, MNRAS, 295, 197

Hernquist, L. 1990, ApJ, 356, 359

Hill, J. M. \& Oegerle, W. R. 1998, AJ, 116, 1529

Jones, C., \& Forman, W. 1984, ApJ, 276, 38

Kauffmann, G., Colberg, J. M.,  Diaferio, A., White, S. D. M. 1999,
MNRAS, 303, 188

Mahdavi, A., Geller, M. J., B\"ohringer, H., \& Ramella, M. 1999,
ApJ, 518, 69

Markevitch, M., Forman, W. R., Sarazin, C. L., \& Vikhlinin, A. 1998,
ApJ, 503, 77

Merritt, D. 1985, AJ, 90, 1027

Metzler, C. A., White, M., Michael, N., \& Loken, C. 1999, ApJ, 520, L9
	
Mohr, J. J., Mathiesen, B., \& Evrard, A. E. 1999, ApJ, 517, 627

Navarro, J. F., Frenk. C. S., \& White, S. D. M. 1997, ApJ, 490, 493

Nulsen, P. E. J. 1986, MNRAS, 221, 377

Peres, C. B., Fabian, A. C., Edge, A. C., Allen, S. W., Johnstone,
R. M., \& White, D. A. 1998, MNRAS, 298, 416

Sarazin, C. L. 1988, X-ray Emission from Clusters of Galaxies
(Cambridge: Cambridge University Press)

Soucail, G., Ota, N., B\"ohringer, H., Czoske, O., Hattori, M., \&
Mellier, Y. 2000, A\&A in press (astro-ph/9911062)

Thomas, P. A. 1998, MNRAS, 299, 349

Tyson, J. A., Kochanski, G. P., \& Dell'Antonio, I. P. 1998, ApJ, 498,
L107

van der Marel, R. P., Magorrian, J., Carlberg, R. G., Yee, H. K. C., 
\& Ellingson, E. 2000, AJ, 119, 2038

Xue Y. \& Wu, X. 2000, ApJ, in press

Zimmermann, H. U., Belloni, T., Izzo, C., Kahabka, P., \& Schwentker,
O. 1993, in ASP Conf. Ser. 52: Astronomical Data Analysis Software and
Systems II, p. 53

\end{references}
\end{document}